\documentstyle[12pt]{article}
              
\title{Quantum Reference Frames and Quantum Transformations.}
\author{M. Toller \\ 
Dipartimento di Fisica dell'Universit\`a, Trento  \\
I.N.F.N. gruppo collegato di Trento, Italia}
 
\newtheorem{proposition}{Proposition}
\newtheorem{assumption}{Assumption}
\begin{document} 
\maketitle  
                 
\begin{abstract}
A quantum frame is defined by a material object subject to the laws of quantum mechanics. The present paper studies the relations between quantum frames, which in the classical case are described by elements of the Poincar\'e group. The possibility of using a suitable quantum group is examined, but some arguments are given which show that a different mathematical structure is necessary.  Some simple examples in lower dimensional spacetimes are treated. They indicate the necessity of taking into account some ``internal'' degrees of freedom of the quantum frames, that can be disregarded in a classical treatment.
\end{abstract} 

\section{Introduction.}

It has been stressed by several authors \cite{Toller1, AK, Rovelli1, Rovelli2} that, from the physical point of view, a frame of reference is defined by a material object of the same nature as the objects that form the system under investigation and the measuring instruments. Then, in principle, one should take into account: 
\begin{enumerate}
\item The gravitational field generated by this object.
\item Its quantum properties that do not permit an
exact determination of its position and of its velocity. 
\end{enumerate}
The gravitational field can be disregarded if the mass of the object that
defines the reference frame is sufficiently small and the quantum effects
are not important if the mass is sufficiently large.  If, in a given 
physical situation, one can choose a mass that satisfies both these 
conditions, the material nature of the reference frames becomes  irrelevant.
Otherwise, we fall into the domain of quantum gravity and the usual 
geometric concepts are not valid any more \cite{Mead, Ferretti, Garay, DFR}. 

In the present work, as in ref.\ \cite{AK}, we disregard the gravitational interaction and we consider quantum reference frames with finite mass $M$. Of course, the physical phenomena can be described completely by considering the limit $M \to \infty$. However, we think that a good understanding of this problem for finite $M$ is preliminary to a treatment taking into account the effects of quantum gravity. 

The relations between classical reference frames are described by elements of the Poincar\'e or of the Galilei group $\cal G$. Our aim is to find a mathematical structure that describes the relations between quantum frames. We start from the remark that in the absence of a classical reference frame the observable quantities, which concern both the objects that define a quantum reference frame and the objects described by the theory, have to be invariant under the action of the group $\cal G$.

Various authors \cite{Majid,GKMMK,LNR,Maggiore} have suggested that the quantum aspects of spacetime can be described by a quantum group \cite{Abe,RTF} obtained from a deformation of the commutative Hopf algebra of the functions defined on the  group $\cal G$. From the point of view of the quantum frames, even in the absence of gravitation, there are two problems:
\begin{enumerate}
\item A quantum group cannot describe the ``internal'' degrees of freedom of the quantum frames and we shall see that they play an essentail role.
\item If we have three quantum frames $F_1$ $F_2$ and $F_3$, the observables which describe the relation between $F_1$ and $F_2$ cannot be compatible (in the sense of quantum theory) with the observables which describe the relations between $F_2$ and $F_3$. This fact cannot be taken into account by a quantum group.
\end{enumerate}
The second problem is an aspect of the ``paradox of quantum frames'' discussed in ref.\ \cite{AK}.

In Section 2 we introduce a general mathematical structure which describes the relations between quantum frames. In Section 3 we consider the classical (non quantum) case and we discuss the connection with the formalism based on Hopf algebras. In Sections 4, 5 and 6 we study some simple examples in zero and one space dimensions, both relativistic and non relativistic. The role of the ``internal'' degrees of freedom of the quantum frames is clarified by these examples.   

It has been suggested in ref.\ \cite{DFR} that in a quantum description of spacetime it is necessary to introduce some new degrees of freedom, besides the spacetime coordinates. We suggest that they can be interpreted as ``internal'' degrees of freedom of the quantum frames.

\section{Relations between Quantum Frames.}

We consider in a flat spacetime a system composed of $n$ objects $F_0,\ldots, F_{n-1}$. Some of them define reference frames and the others are the objects described by the theory. In order to simplify the problem, we assume that these objects do not interact mutually, but only with the measuring instruments. This assumption is justified only if we can disregard
gravitation, a long range interaction which involves in an universal way all the kinds of physical objects. 

In a quantum treatment, the states of the system are 
described by the Hilbert space
\begin{equation} 
{\cal H}^{(n)} = {\cal H}_0 \otimes  {\cal H}_1 \otimes \cdots  
\otimes {\cal H}_{n-1}.
\end{equation}
We assume that the bounded observables are described by Hermitean operators belonging to the algebra\footnote{In a more refined treatment, one should consider a smaller $C^*$ algebra  ${\cal A}^{(n)} \subset {\cal L}({\cal H}^{(n)})$, in order to take into account the possible occurrence of superselection rules and the topological properties of the system \cite{Giles}.} ${\cal A}^{(n)} = {\cal L}({\cal H}^{(n)})$ of the bounded operators in ${\cal H}^{(n)}$. The observables which concern the object $F_k$ are described by the algebra ${\cal A}_k = {\cal L}({\cal H}_k)$ and the observables which concern the pair of objects $\{F_i, F_k\}$  are described by the algebra ${\cal A}_{ik} = {\cal L} ({\cal H}_i \otimes {\cal H}_k)$. We define the injective homomorphisms $j^{(n)}_k: {\cal A}_k \to {\cal A}^{(n)}$ and $j^{(n)}_{ik}: {\cal A}_{ik} \to {\cal A}^{(n)}$ by means of the formulas
\begin{equation} 
j^{(n)}_k(A_k) = B_0 \otimes \cdots \otimes B_{n-1}, \qquad B_k = A_k \in {\cal A}_k, \qquad B_j = 1 \quad {\rm for} \quad j \neq k,
\end{equation}  
\begin{displaymath} 
j^{(n)}_{ik}(A_i \otimes A_k) = B_0 \otimes \cdots \otimes B_{n-1}, \qquad  B_i = A_i \in {\cal A}_i, \qquad B_k = A_k \in {\cal A}_k, 
\end{displaymath}
\begin{equation} 
  B_j = 1 \quad {\rm for} \quad j \neq i, \quad j \neq k.
\end{equation}  
These homomorphisms formalize the concept of subsystem, which is essential for the following considerations.
                                                  
We adopt the Heisenberg picture for time evolution. An object in a given 
Heisenberg state defines a four-dimensional reference frame and not a set
of three-dimensional frames depending on time. In particular, the object defines the origin of the time scale with the precision permitted by the  indeterminacy relations.

We indicate by $\cal G$ the Poincar\'e or the Galilei group, or some other
spacetime symmetry group. It acts on ${\cal H}^{(n)}$ by means of the unitary 
representation  
\begin{equation} 
 g \to U^{(n)}(g) = U_0(g) \otimes  U_1(g) \otimes \cdots  \otimes  U_{n-1}(g),  \qquad g \in {\cal G}. 
\end{equation}    
The description of the system by means of the algebra ${\cal A}^{(n)}$ implicitly assumes the existence of an ``external'' object with a very large mass which defines a classical reference frame. If a classical reference 
frame is not available, only the $\cal G$-invariant elements of ${\cal A}^{(n)}$ represent observables. In other words, $\cal G$ has to be considered as a global gauge group\footnote{We do not consider a local Poincar\'e gauge invariance, which would lead to a theory of gravitation \cite{Hehl}.}.  We indicate by ${\cal B}^{(n)}$ the subalgebra of ${\cal A}^{(n)}$ composed of the $\cal G$-invariant elements, namely the elements which commute with all the operators $U^{(n)}(g)$. It is a von Neumann algebra \cite{Dixmier}, namely it is symmetric (self-adjoint) and closed under the weak operator topology.
        
We also define the algebra ${\cal B}_k$ of the $\cal G$-invariant 
elements of ${\cal A}_k $ which describes the ``internal'' degrees of freedom of the object $F_k$ and the algebra ${\cal B}_{ik}$ of the invariant elements of ${\cal A}_{ik}$ which describes, besides the ``internal'' degrees of freedom of the objects $F_i$ and $F_k$, the ``relative'' parameters, which define the relative position, time and velocity of the two objects. These relative parameters are the quantum analogs of the parameters defining the element of $\cal G$ which connects two classical reference frames. The homomorphisms $j^{(n)}_k$ and $j^{(n)}_{ik}$ can be restricted to these invariant subalgebras.  We put  
\begin{equation}
j^{(n)}_k {\cal B}_k = {\cal B}^{(n)}_k \subset {\cal B}^{(n)}, \qquad j^{(n)}_{ik} {\cal B}_{ik} = {\cal B}^{(n)}_{ik} = {\cal B}^{(n)}_{ki} \subset {\cal B}^{(n)}.
\end{equation}  
If $i, j, k,$ are different indices, we have
\begin{equation} \label{Intersection}
{\cal B}^{(n)}_{ji} \cap {\cal B}^{(n)}_{jk} = {\cal B}^{(n)}_j,     \qquad n \geq 3.
\end{equation}  

Not all the kinds of objects are suitable for the definition of a reference frame. For instance, a spherically symmetric object cannot specify the directions of the spatial coordinate axes. If the object $F_j$ defines a reference frame accurately, we expect that the algebra ${\cal B}^{(n)}_{jk}$ describes the object $F_k$ completely and that the following property holds:
\begin{assumption}  \label{Frame}
The algebra ${\cal B}^{(n)}$ is generated by the $n-1$ subalgebras ${\cal B}^{(n)}_{jk}$ with $j$ fixed and $k \neq j$.
\end{assumption} 
In the following Sections we shall test this assumption in some simple models. Without any loss of generality, we can assume that $j = 0$.

We remember \cite{Dixmier} that the von Neumann algebra generated by a symmetric (self-adjoint) set $\cal S$ of operators is the double commutant ${\cal S}''$. If we indicate by ${\cal U}^{(n)}$ the von Neumann algebra generated by all the operators $U^{(n)}(g)$, we have  
\begin{equation} 
{\cal B}^{(n)} = {\cal U}^{(n)\prime}.
\end{equation}  
It is often convenient to describe a quantum system in terms of unbounded observables, which correspond to unbounded self-adjoint operators. If   $\cal S$ is a set of self-adjoint not necessarily bounded operators, we indicate by ${\cal S}'$ the algebra of the bounded operators which commute with all the spectral projectors of all the operators in $\cal S$. We say that a self-adjoint operator is associated to a von Neumann algebra if its spectral projectors belong to the algebra. 

If we indicate by ${\cal S}_{ik}$ a set of self-adjoint operators associated to ${\cal B}^{(n)}_{ik}$ and we have
\begin{equation} \label{Condition}
{\cal S}_0' = {\cal U}^{(n)}, \qquad  
{\cal S}_0 = \bigcup_{k=1}^{n-1} {\cal S}_{0k}, 
\end{equation} 
we obtain
\begin{equation} 
{\cal S}_0'' = {\cal B}^{(n)}
\end{equation}       
and the Assumption \ref{Frame}
is satisfied for the frame $F_0$.  The same reasoning for $n = 2$ shows that ${\cal B}^{(2)} = {\cal B}_{01}$ is generated by ${\cal S}_{01}$ if ${\cal S}'_{01} = {\cal U}^{(2)}$.  
 
We say that two subalgebras commute if all the elements of the first subalgebra commute with all the elements of the second one. One can easily see that if the indices $i, k, j, l$ are all different, the subalgebras ${\cal B}^{(n)}_{ik}$ and ${\cal B}^{(n)}_{jl}$ commute. However in a quantum theory one cannot assume that the subalgebras ${\cal B}^{(n)}_{jk}$ and ${\cal B}^{(n)}_{jl}$ commute.  In fact we have:  
\begin{proposition} \label{Paradox} 
If the objects $F_0$, $F_1$ and $F_2$ define quantum frames, namely they satisfy the Assumption \ref{Frame}, and all the pairs of the subalgebras ${\cal B}^{(3)}_{01}$, ${\cal B}^{(3)}_{02}$ and ${\cal B}^{(3)}_{12}$ commute, then the algebra ${\cal B}^{(3)}$ is commutative and describes a classical system.
\end{proposition}
In fact, since the subalgebras ${\cal B}^{(3)}_{01}$ and ${\cal B}^{(3)}_{02}$ generate ${\cal B}^{(3)}$, ${\cal B}^{(3)}_{12}$ commutes with ${\cal B}^{(3)}$. In a similar way we see that ${\cal B}^{(3)}_{02}$ commmutes with ${\cal B}^{(3)}$. A third application of the Assumption \ref{Frame} shows that ${\cal B}^{(3)}$ commutes with itself, namely it is commutative.

Our aim is to study the relations between quantum frames and therefore we consider a set of $n$ objects all of the same kind\footnote{We assume, however, that they have some different quantum number in order to avoid the symmetrization or the antisymmetrization of the wave function.} which describe $n$ quantum frames and therefore satisfy the Assumption \ref{Frame}. We can simplify the notations by means of the identifications ${\cal B}_k = {\cal B}^{(1)}$ and ${\cal B}_{ik} = {\cal B}^{(2)}$. In order to obtain a compact notation for the various homomorphisms of the algebras ${\cal B}^{(n)}$, we introduce the sets $\Delta_n = \{0,\ldots, n-1 \}$ and for $n \geq m \geq 1$ the sets $G_{nm}$ of the injective functions $\pi: \Delta_m \to \Delta_n$.  If $\pi \in G_{nm}$, we consider the injective homomorphism $j_{\pi}: {\cal A}^{(m)} \to {\cal A}^{(n)}$ defined by
\begin{displaymath} 
j_{\pi}(A_0 \otimes \ldots \otimes A_{m-1}) = B_0 \otimes \ldots \otimes B_{n-1},  
\end{displaymath}
\begin{equation} 
B_{\pi(k)} = A_k, \qquad  
B_j = 1 \quad {\rm for} \quad j \not\in \pi(\Delta_m).
\end{equation}
These homomorphisms can be restricted to the $\cal G$-invariant subalgebras, namely we have $j_{\pi}: {\cal B}^{(m)} \to {\cal B}^{(n)}$. They have the property
\begin{equation} \label{Composition}
j_{\pi} j_{\sigma} = j_{\pi \circ \sigma}, \qquad
\pi \in G_{np}, \qquad  \sigma \in G_{pm}, \qquad n \geq p \geq m \geq 1.
\end{equation}

We have 
\begin{equation} 
j^{(n)}_k = j_{\pi}, \qquad \pi \in G_{n1}, \qquad \pi(0) = k,
\end{equation}
\begin{equation} 
j^{(n)}_{ik} = j_{\pi}, \qquad \pi \in G_{n2}, \qquad 
\pi(0) = i, \quad \pi(1) = k.
\end{equation}
We introduce  the automorphism $s$ of the algebra ${\cal B}^{(2)}$ by means of the formula
\begin{equation} 
s = j_{\pi}, \qquad \pi \in G_{2,2}, \qquad \pi(0) = 1, \qquad \pi(1) = 0.
\end{equation}
and from eq.\ (\ref{Composition}) we get
\begin{equation}  \label{Exchange}
j^{(n)}_{ki} s =j^{(n)}_{ik}, \qquad  s^2 = 1.
\end{equation}
       
The case $n = 3$ is particularly interesting: we have an algebra ${\cal B}^{(3)}$ that contains three isomorphic subalgebras ${\cal B}^{(3)}_{01}$, ${\cal B}^{(3)}_{02}$ and ${\cal B}^{(3)}_{12}$, and is generated by any pair of them. This means, for instance, that an element of ${\cal B}^{(3)}_{12}$, which concerns the relation between the frames $F_1$ and $F_2$, can be expressed as a weak limit of algebraic expressions containing elements of ${\cal B}^{(3)}_{01}$ and ${\cal B}^{(3)}_{02}$. In other words,  the mathematical structure of ${\cal B}^{(3)}$ with its three subalgebras describes the composition law of the transformations of quantum frames, which can be considered as the quantum version of the multiplication law of the group $\cal G$. It is not yet clear if the algebras ${\cal B}^{(n)}$ with $n > 3$ contain further physically relevant informations. One can remark that the formulation of the associative law requires three transformations and four reference frames.

In the treatment given above the algebras  ${\cal B}^{(n)}$ and the homomorphisms $j_{\pi}$ have been constructed starting from the group $\cal G$ and its unitary representations. One can conceive a more general kind of theory in which the group $\cal G$ does not exist and the relations between quantum reference frames are described directly by the von Neumann algebras ${\cal B}^{(n)}$ and by the homomorphisms $j_{\pi}$. Of course, some properties which follow from the construction based on the group $\cal G$ have to be assumed explicitly. Besides the Assumption \ref{Frame}, we require:
\begin{assumption}  \label{Compos}
The homomorphisms $j_{\pi}$ are injective and satisfy eq.\ (\ref{Composition}). 
\end{assumption} 
\begin{assumption}  \label{Commut}
If the indices $i, k, j, l$ are all different, the subalgebras ${\cal B}^{(n)}_{ik}$ and ${\cal B}^{(n)}_{jl}$ commute.
\end{assumption}
One can also assume that eq.\ (\ref{Intersection}) is valid, possibly after a redefinition of the algebra  ${\cal B}^{(1)}$. Note that our assumptions concern only the algebras, without any reference to the underlying Hilbert spaces.

A special case of eq.\ (\ref{Composition}) is
\begin{equation} \label{Compo}
j_{\pi} j^{(m)}_{ik} = j^{(n)}_{\pi(i), \pi(k)}, \qquad \pi \in G_{nm}, \qquad  n \geq m \geq 2
\end{equation}  
and from the Assumption \ref{Frame}, we see that the homomorphism  $j_{\pi}$ is uniquely determined by the homomorphism $j^{(n)}_{ik}$ and $j^{(m)}_{ik}$. Note however that a homomorphism $j_{\pi}$ that satisfies eq. (\ref{Compo}) does not necessarily exist. 

From our assumptions one can derive the following result, that is stronger than Proposition \ref{Paradox}: 
\begin{proposition} \label{Paradox1} 
If the indices $i, k, j$ are all different and the subalgebras ${\cal B}^{(n)}_{jk}$ and ${\cal B}^{(n)}_{ji}$ commute, then all the algebras  ${\cal B}^{(m)}$ with  $m \leq n$ are commutative.
\end{proposition} 
The automorphisms $j_{\pi}$ with $\pi \in G_{nn}$ permute the subalgebras ${\cal B}^{(n)}_{jk}$ and, taking into account the Assumption \ref{Commut}, we see that all the pairs of subalgebras ${\cal B}^{(n)}_{ik}$ commute. As in the proof of Proposition \ref{Paradox}, it follows that all the subalgebras ${\cal B}^{(n)}_{jk}$ belong to the centre of ${\cal B}^{(n)}$ and that ${\cal B}^{(n)}$ is commutative. The algebras ${\cal B}^{(m)}$ with $m < n$ are isomorphic to subalgebras of ${\cal B}^{(n)}$ and the Proposition is proven.  

Proposition \ref{Paradox1} means that in a quantum theory the observables which describe a transformation between the frames $F_j$ and $F_i$ cannot be compatible with the observables which describe a transformation between the frames $F_j$ and $F_k$. This fact is strictly related to the ``paradox of quantum frames'' discussed in ref.\ \cite{AK}. 

\section{Classical reference frames.} 

By considering the classical limit of the formalism developed in the preceding section, we clarify the meaning of various concepts and we can understand why the relations between quantum frames cannot be described by a quantum group \cite{Abe,RTF} obtained by a ``quantization'' of the group $\cal G$.

In a classical theory the algebras ${\cal A}_0,\ldots, {\cal A}_{n-1}$ and ${\cal A}^{(n)}$ are commutative and can be considered as algebras of $L^{\infty}$ functions (essentially bounded measurable functions) on the phase spaces $\Gamma_0,\ldots, \Gamma_{n-1}$ and
\begin{equation} 
\Gamma^{(n)} = \Gamma_0 \times \Gamma_1 \times \cdots \times \Gamma_{n-1}, 
\end{equation}
endowed with their Liouville measures. If we adopt the weak topology defined by the duality with the spaces $L^1$, we are dealing with commutative von Neumann algebras of operators acting multiplicatively on the corresponding $L^2$ function spaces. The Lie group $\cal G$ acts smoothly on the phase spaces preserving the Liouville measures. For this action we use the notation $(g, x_k) \to gx_k, \quad g \in {\cal G}, \quad x_k \in \Gamma_k$. A function $f \in {\cal B}^{(n)}$ has the invariance property
\begin{equation}  \label{Invariance}
f(gx_0,\ldots, gx_{n-1}) =  f(x_0,\ldots, x_{n-1}).
\end{equation}  

We say that $\cal G$ acts freely on $\Gamma_0$ if, for any $x_0 \in \Gamma_0$, $\,\,gx_0 = x_0$ implies $g=e$ ($e$ is the unit element). We also say that two submanifolds intersect transversely at a point if the vectors tangent to the submanifolds at this point generate the whole tangent space. Then we have:
\begin{proposition}
If $\cal G$ acts freely on $\Gamma_0$ and there is a submanifold $\hat\Gamma_0 \subset \Gamma_0$ that intersects transversely every orbit of $\Gamma_0$ in one point, ${\cal B}^{(n)}$ is generated by the $n-1$ subalgebras ${\cal B}^{(n)}_{0k}$, $(k = 1,\ldots, n-1)$. Under these conditions the Assumption \ref{Frame} is satisfied by the frame $F_0$.
\end{proposition}      
This result could be obtained under much weaker hypotheses, but our purpose is only to give an example.

In order to sketch a proof, we remark that every $x_0 \in \Gamma_0$ can be written uniquely in the form $x_0 = g y_0$, with $ y_0 \in \hat\Gamma_0$, and this relation defines a diffeomorphism of $\Gamma_0$ and ${\cal G} \times \hat\Gamma_0$. The Liouville measure on $\Gamma_0$ is transformed into the product of a left invariant measure on $\cal G$ and a suitable measure on $\hat\Gamma_0$. Since we have
\begin{equation} 
f(g y_0, x_1,\ldots, x_n) = f(y_0, g^{-1} x_1,\ldots, g^{-1} x_{n-1}), \qquad f \in {\cal B}^{(n)},
\end{equation}              
we can identify the algebra ${\cal B}^{(n)}$ with the algebra of the $L^{\infty}$ functions defined on $\hat\Gamma_0 \times \Gamma_1 \times \cdots \times \Gamma_{n-1}$. The subalgebra generated by the subalgebras ${\cal B}^{(n)}_{0k}$ contains all the products of $L^{\infty}$ functions of the kind $f(y_0, x_k)$ and their linear combinations. This set of functions is dense in ${\cal B}^{(n)}$.

We consider the special case in which all the objects are of the same kind and we put $\Gamma_0 = \ldots = \Gamma_{n-1} = \Gamma$. We also assume that $\cal G$ acts freely and transitively on $\Gamma$. This means that the objects have no ``internal'' degree of freedom and $\cal G$ and $\Gamma$ are diffeomorphic. If we choose a point $y$ of $\Gamma$, and we put  
\begin{equation} 
f(g_0 y, g_1 y,\ldots, g_{n-1} y) = \hat f(g_0^{-1} g_1,\ldots, g_0^{-1} g_{n-1}), 
\end{equation}      
we can consider ${\cal B}^{(n)}$ as a space of functions of $n-1$ arguments in $\cal G$. In particular, ${\cal B}^{(2)}$ is a space of functions on $\cal G$ and we can write
\begin{equation} 
{\cal B}^{(n)} = {\cal B}^{(2)} \otimes \cdots \otimes {\cal B}^{(2)}, \qquad (n-1 \,\,\, {\rm factors}).
\end{equation} 
    
 If we restrict our attention to continuous functions, ${\cal B}^{(2)}$ has a structure of Hopf algebra \cite{Abe,RTF}. If we want to work with $L^{\infty}$ functions, we have to renounce to the existence of a counit. We have
\begin{equation} 
[j_{\pi}f](x_0, x_1,\ldots, x_{n-1}) = f(x_{\pi(0)}, x_{\pi(1)},\ldots, x_{\pi(m-1)}), \qquad  \pi \in G_{nm},
\end{equation}   
\begin{equation} 
[j_{\pi}\hat f](g_1,\ldots, g_{n-1}) = \hat f(g_{\pi(0)}^{-1} g_{\pi(1)},\ldots, g_{\pi(0)}^{-1} g_{\pi(m-1)}), 
\end{equation} 
where in the right hand side we have to put $g_0 = e$. 
We consider some particular cases:
\begin{equation}  
[s \hat f](g) = \hat f(g^{-1}), \qquad \hat f \in {\cal B}^{(2)}, 
\end{equation}                            
namely the automorphism $s$ is the antipode of the Hopf algebra ${\cal B}^{(2)}$; 
\begin{equation}
[j^{(3)}_{12} \hat f](g_1, g_2) = \hat f(g_1^{-1} g_2),
\end{equation}
namely
\begin{equation} \label{Coproduct}
j^{(3)}_{12} = (s \otimes 1) C,
\end{equation}
where $C: {\cal B}^{(2)} \to {\cal B}^{(2)} \otimes {\cal B}^{(2)} = {\cal B}^{(3)}$ is the coproduct of the Hopf algebra. From the formula
\begin{equation}  
[j^{(n)}_{0k} \hat f](g_1,\ldots, g_{n-1}) = \hat f(g_k),  
\end{equation}  
we obtain
\begin{equation} \label{Homo}
j^{(3)}_{01} A = A \otimes 1, \qquad         
j^{(3)}_{02} A = 1 \otimes A, \qquad  A \in  {\cal B}^{(2)}.
\end{equation}

The formulas (\ref{Coproduct}) and (\ref{Homo}) seem very natural in a formalism based on a Hopf algebra, but their generalization to non-commutative Hopf algebras (quantum groups) is not possible. In fact, we see from eq.\ (\ref{Homo}) that the subalgebras ${\cal B}^{(3)}_{01}$ and ${\cal B}^{(3)}_{02}$ commute and it follows from the Proposition  \ref{Paradox1} that ${\cal B}^{(3)}$ and ${\cal B}^{(2)}$ must be commutative. Moreover, from eqs.\ (\ref{Homo}) and  (\ref{Intersection}) we have that ${\cal B}^{(3)}_{0}$ and therefore ${\cal B}^{(1)}$ contain only the multiples of the unit element. We see that a quantum group cannot describe the ``internal'' degrees of freedom of the quantum frames and the quantum aspects which are required by Proposition \ref{Paradox1}.    

\section{The simplest example.}
   
As a first example of the general formalism, we consider the case in which $\cal G$ contains only the time translations and the objects that define the frames are essentially clocks. The algebra ${\cal B}^{(n)}$ is composed of the elements of ${\cal A}^{(n)}$ which commute with the Hamiltonian. Time measurements and clocks in quantum mechanics have been discussed by several authors \cite{Pauli, Armstrong, SW, EF, AB, Rosenbaum, ORG, Rovelli3}.

The simplest realization of a clock is a system with one degree of freedom. We consider a set of $n$ clocks of this kind and we describe the clock $F_k$ by means of the canonical variables $q_k, p_k$ with ($\hbar = 1$). We have
\begin{equation} \label{Canonical}
[q_i, p_k] = i \delta_{ik},
\end{equation} 
\begin{equation}
 H = \sum_k H_k, \qquad  H_k = 2^{-1} p_k^2.
\end{equation} 
This system can be reinterpreted as a single free particle in an $n$-dimensional space. The momenta $p_k$ and the components of the ``angular momentum'' 
\begin{equation}  \label{Angular}
L_{ik} = q_i p_k - q_k p_i
\end{equation} 
are conserved self-adjoint operators. They satisfy the commutation relations
(\ref{Canonical}) and
\begin{equation}  \label{Commutation}
[L_{ik}, p_j] = i \delta_{ij} p_k  - i \delta_{kj} p_i, 
\end{equation} 
\begin{equation}
[L_{ik}, L_{rs}] = i \delta_{ir} L_{ks} - i \delta_{is} L_{kr}  
- i \delta_{kr} L_{is} + i \delta_{ks} L_{ir}.
\end{equation}
                   
With the notation introduced in Sect.\ 2, we put ${\cal S}_{jk} = \{p_j, p_k, L_{jk}\}$. The set ${\cal S}_0$ contains all the momenta $p_k$ and the quantities $L_{0k}$ with $k = 1,\ldots, n-1$, which generate the whole rotation group. It follows that the commutant ${\cal S}_0'$ is composed of the rotation-invariant bounded functions of the momenta $p_k$, namely the bounded functions of the Hamiltonian $H$, which form, in the present case, the algebra ${\cal U}^{(n)}$.  We have seen that the condition (\ref{Condition}) is satisfied and the Assumption \ref{Frame} is valid. 

The observable $q_k$ can be interpreted as the running time directly read on the clock, while $p_k$ is the rate of the clock. In the classical case, the true time is given by the ratio $T_k = q_k p_k^{-1}$. In the quantum case one can define in a suitable linear subspace of $\cal H$ the Hermitean operator 
\begin{equation} \label{Time}
T_k = (2p_k)^{-1} q_k + q_k (2p_k)^{-1},
\end{equation}
which satisfies the commutation relation 
\begin{equation} 
i [H_k, T_k] = 1, 
\end{equation} 
or, more exactly,
\begin{equation}
\exp(it H_k) T_k \exp(-it  H_k) =  T_k + t. 
\end{equation} 

An argument due to Pauli \cite{Pauli} shows that under these conditions
$T_k$ cannot be self-adjoint. In fact, if it is self-adjoint one can prove that
\begin{equation}
\exp(i E T_k) H_k \exp(-i  E T_k) = H_k - E
\end{equation}          
and this equation contradicts the fact that $H_k$ has a spectrum bounded from below.  Note that if an operator is not self-adjoint, it has no spectral representation, one cannot define bounded functions of it and it cannot represent an observable which has an arbitrarily small dispersion on a complete set of states.
                           
The relative time of two clocks is given by
\begin{equation} 
T_i - T_k = (2 p_i p_k)^{-1} L_{ik} + L_{ik} (2 p_i p_k)^{-1}.
\end{equation}
From the commutation relations (\ref{Commutation}), we see that if we fix $p_i$ or $p_k$, the quantity $L_{ik}$ and the relative time are completely undetermined. We see that the ``internal'' variables $p_i$ play an important role and cannot be eliminated by imposing that they take a fixed value. The same remark holds for the more complicated models described in the following Sections.  

\section{One-dimensional non-relativistic frames.}

Now we assume that $\cal G$ is the Galilei group in one space dimension. An object which identifies a reference frame must contain al least two particles. We consider $n$ objects composed of two free particles and
we indicate by $M_k$ the total mass and by $\mu_k$ the reduced mass of the object $F_k$. We adopt as canonical variables the centre of mass coordinate $X_k$, the total momentum $P_k$, the relative coordinate $\mu_k^{-1/2} q_k$ and the variable  $p_k$ conjugated to $q_k$. The Hamiltonian is 
\begin{equation}
H = \sum_k H_k, \qquad H_k =  (2 M_k)^{-1} P_k^2 + 2^{-1} p_k^2.
\end{equation}
The canonical variables $q_k$ and $p_k$ have been scaled in order to eliminate the reduced mass $\mu_k$. They describe a clock as in the preceding section. 

The generators of the time and space translations are $H$ and the total momentum
\begin{equation}
P = \sum_k P_k,
\end{equation} 
while the Galilei boosts are generated by
\begin{equation}
K = -\sum_k M_k X_k.
\end{equation}
Note that
\begin{equation}
[H, P] = 0, \qquad  [H, K] = i P, \qquad  [P, K] = i \sum_k M_k.
\end{equation}       

The $\cal G$-invariant variables are the internal variables $p_k$, the relative variables $L_{ik}$ discussed in the preceding section, the relative velocities of the centres of mass 
\begin{equation}
V_{ik} = M_k^{-1} P_k - M_i^{-1} P_i,
\end{equation}
and the relative quantities
\begin{equation}
Y_{ik} =  p_k (X_k - X_i) -  q_k V_{ik} .
\end{equation}
The quantity
\begin{equation}
(2p_k)^{-1}  Y_{ik} + Y_{ik} (2p_k)^{-1} = X_k - X_i - T_k  V_{ik}
\end{equation} 
can be interpreted as the space coordinate of the origin of the frame $F_k$ measured with respect to the frame $F_i$.

We treat the Hilbert space ${\cal H}^{(n)}$ as a space of $L^2$ functions defined on the $2n$-dimensional space $\cal P$ of the variables $p_k$ and $P_k$ (we use the same symbol for these variables and for the corresponding multiplication operators). The operators $p_k$ and $V_{ik}$ are clearly self-adjoint. The operators $L_{ik}$ and $Y_{ik}$ are generators of a group of linear measure-preserving transformations of the space $\cal P$ and they can be considered in a natural way as self-adjoint.  One can see by means of a direct calculation that, if one excludes the lower dimensional manifold where $p_k = 0$ and $V_{ik} = 0$, the orbits of this group have codimension two and are just the manifolds on which the quantities $H$ and $P$ take constant values.

We put ${\cal S}_{ik} = \{p_i, p_k, L_{ik}, V_{ik}, Y_{ik}\}$ and we see that ${\cal S}_0$ contains all the ``internal'' variables $p_k$ and the relative variables $L_{0k}$, $V_{0k}$ and $Y_{0k}$. We replace in the space $\cal P$ the coordinates $P_k$ by their linear combinations  $P$ and $V_{0k}$. If we indicate by ${\cal H}'$ the space of the $L^2$ functions of the variable $P$ and by ${\cal H}''$ the space of the $L^2$ functions of the other variables $q_k$ and $V_{0k}$, we have ${\cal H}^{(n)} = {\cal H}' \otimes {\cal H}''$. An operator  $A \in {\cal S}_0'$ is diagonal in the variables $p_k$ and $V_{0k}$, but acts in a general way on the variable $P$. In other words, we have a direct integral decomposition \cite{Dixmier} of $A$ into operators $A(p_k, V_{0k}) \in {\cal W}$, where $\cal W$ is the algebra of the bounded operators acting on ${\cal H}'$. Since $A$ must commute with  $L_{0k}$ and $Y_{0k}$, $A(p_k, V_{0k})$ must be constant on the orbits of the linear group generated by these operators, namely it depends only on the variable
\begin{equation}
\hat H = H - (2 \sum_k M_k)^{-1} P^2 =   2^{-1} \sum_k p_k^2 
+ (4 \sum_k M_k)^{-1} \sum_{ij} M_i M_j V_{ij}^2.               
\end{equation} 
   
In conclusion, the elements of ${\cal S}_0'$ are given as direct integrals of bounded measurable functions $A(\hat H)$ with values in $\cal W$ and we can write \cite{Dixmier} ${\cal S}_j' = {\cal W} \otimes {\cal V}$, where ${\cal V}$ is the algebra of the bounded functions of $\hat H$ considered as operators on ${\cal H}''$. Note that the algebra $\cal W$ is generated by the operators $P$ and 
\begin{equation}
K = i \sum_k M_k \frac{\partial}{\partial P}.
\end{equation} 
It follows that ${\cal S}_0'$ is generated by the operators $P$, $K$ and $\hat H$, namely it coincides with ${\cal U}^{(n)}$. Also in this case the condition (\ref{Condition}) is satisfied and the Assumption \ref{Frame} is valid.     

\section{One-dimensional relativistic frames.}

As in the non relativistic case, we consider $n$ frames, each one defined by means of two free particles, described by two irreducible unitary representations of the Poincar\'e group with masses $m_k'$ and $m_k''$. The generators of the Poincar\'e transformations are given by
\begin{equation}
H = \sum_k H_k, \qquad P = \sum_k P_k, \qquad K = \sum_k K_k,
\end{equation}
where (assuming $c = 1$)
\begin{displaymath}
H_k = H_k' + H_k'' = 
((P_k')^2 + (m_k')^2)^{1/2} + ((P_k'')^2 + (m_k'')^2)^{1/2}, 
\end{displaymath}
\begin{equation}
 P_k = P_k' + P_k'', \qquad K_k = K_k' + K_k''. 
\end{equation}
The commutation relations are
\begin{equation}
[H, P] = 0, \qquad  [H, K] = i P, \qquad  [P, K] = i H
\end{equation}  
and similar relations hold for the generators acting on the various subsystems.

It is convenient to describe the object $F_k$ by means of the variables $H_k$, $P_k$, $K_k$ and
\begin{equation}
p_k = P_k' H_k'' - P_k'' H_k', \qquad
q_k = H_k' K_k'' - H_k'' K_k', \qquad
r_k = K_k' P_k'' - K_k'' P_k',
\end{equation}                
which are connected by the identities
\begin{equation}
H_k^2 - P_k^2 =  (m_k')^2 + (m_k'')^2  + 2 ((m_k' m_k'')^2 + p_k^2)^{1/2}
= M_k^2,
\end{equation} 
\begin{equation}
p_k K_k + q_k P_k + r_k H_k = 0.
\end{equation} 
We have the commutation relations
\begin{displaymath}
[H, p_k] = 0, \qquad  [H, q_k] = -i p_k, \qquad   [H, r_k] = 0, 
\end{displaymath} 
\begin{displaymath}
[P, p_k] = 0, \qquad  [P, q_k] = 0, \qquad   [P, r_k] = -i p_k,
\end{displaymath}
\begin{equation}
[K, p_k] = 0, \qquad  [K, q_k] = i r_k, \qquad   [K, r_k] = i q_k.
\end{equation}    
\begin{displaymath}
[p_k, q_k] = -i H_k ((m_k' m_k'')^2 + p_k^2)^{1/2}, 
\end{displaymath}    
\begin{displaymath} 
[r_k, p_k] = -i P_k ((m_k' m_k'')^2 + p_k^2)^{1/2}, 
\end{displaymath} 
\begin{equation}
[q_k, r_k] = -i K_k ((m_k' m_k'')^2 + p_k^2)^{1/2}.
\end{equation}

These relations and the non-relativistic limit suggest the interpretation of the quantities   
\begin{equation}
\hat T_k = t - T_k,  \qquad 
\hat X_k = (2 p_k)^{-1} r_k + r_k (2 p_k)^{-1},
\end{equation} 
where $T_k$ is given  by eq.\ (\ref{Time}), as the time and space coordinates of the origin of the frame $F_k$ measured in a classical frame. The relative coordinates of the frames $F_k$ and $F_i$ are given by 
\begin{displaymath}
\hat T_k - \hat T_i =
 (2 p_i p_k)^{-1} Q_{ik} + Q_{ik} (2 p_i p_k)^{-1}, 
\end{displaymath} 
\begin{equation}
\hat X_k - \hat X_i = 
(2 p_i p_k)^{-1} R_{ik} + R_{ik} (2 p_i p_k)^{-1},
\end{equation} 
where
\begin{equation}
Q_{ik} = q_i p_k - q_k p_i, \qquad
R_{ik} = p_i r_k - p_k r_i.
\end{equation} 
These quantities
commute with $H$ and $P$  and satisfy the commutation relations  
\begin{equation}
[K, Q_{ik}] = -i R_{ik}, \qquad
[K, R_{ik}] = -i Q_{ik}.
\end{equation} 
It follows that the quantities  
\begin{equation}
L_{ik} = (H_i + H_k) Q_{ik} - (P_i + P_k) R_{ik},
\end{equation}\begin{equation}
Y_{ik} = (H_i + H_k) R_{ik} - (P_i + P_k) Q_{ik} 
\end{equation}
are Poincar\'e invariant. They are proportional to the relative time and space coordinates of the frames $F_i$ and $F_k$ measured in the center of mass frame of these two objects. Note that the symbols which appear in the present section have not exactly the same meaning as in the preceding sections.

Another  Poincar\'e invariant quantity is the relative rapidity of two frames which is given by 
\begin{equation}
V_{ik} = \zeta_k - \zeta_i, \qquad  \zeta_i = \tanh^{-1}\frac{P_i}{H_i}.  
\end{equation} 
               
As in the preceding section, we put ${\cal S}_{ik} = \{p_i, p_k, V_{ik}, L_{ik}, Y_{ik}\}$ and ${\cal S}_0$ contains the operators  $p_k$, $V_{0k}$, $L_{0k}$ and $Y_{0k}$. We describe the vectors of the Hilbert space ${\cal H}^{(n)}$ by means of wave functions in momentum space $\cal P$, square integrable with respect to the Lorentz invariant measure
\begin{equation}
d \mu = \prod_{k=0}^{n-1} \frac{d P'_k}{2 H'_k} \frac{d P''_k}{2 H''_k} =
\prod_{k=0}^{n-1} 2^{-2} ((m_k' m_k'')^2 + p_k^2)^{-1/2} dp_k d\zeta_k.
\end{equation} 
Then $p_k$ and $V_{ik}$ are self-adjoint multiplication operators.
Since we have 
\begin{equation}
K'_k = -i H'_k \frac{\partial}{\partial P'_k}, \qquad
K''_k = -i H''_k \frac{\partial}{\partial P''_k},
\end{equation} 
we see that $L_{ik}$ and $Y_{ik}$ are first order differential operators with smooth coefficients. They generate one-parameter groups of (non linear) measure-preserving diffeomorphism of the compact set in $\cal P$ defined by the inequality $H \leq C$. These diffeomorphisms can be extended to the whole manifold $\cal P$ and they define continuous groups of unitary operators in ${\cal H}^{(n)}$. It follows that the generators $L_{ik}$ and $Y_{ik}$ are self-adjoint. 

A detailed analysis of the group of diffeomorphism generated by $L_{0k}$ and $Y_{0k}$ shows that, if we exclude the lower dimensional manifold where all the variables $p_k$ vanish, the orbits have codimension two and are characterized by constant values of $H$ and $P$. The arguments proceeds in close analogy with the non-relativistic case. We introduce in the space $\cal P$ the coordinates $\zeta_0$, $p_k$ and $V_{0k}$ and we write  ${\cal H}^{(n)} = {\cal H}' \otimes {\cal H}''$, where ${\cal H}'$ contains the $L^2$ functions of $\zeta_0$ and  ${\cal H}''$ contains the $L^2$ functions of the other coordinates. An operator $A \in {\cal S}_0'$ is diagonal in the variables $p_k$ and $V_{0k}$ and can be represented as a direct integral of a function with values in the algebra  ${\cal W}$ of the bounded operators in ${\cal H}'$. This function is constant on the orbits and therefore depends only on the quantity
\begin{equation}
M^2 = H^2 - P^2 = \sum_{ik} M_i M_k \cosh V_{ik}.
\end{equation} 
We have ${\cal S}_0' = {\cal W} \otimes {\cal V}$, where ${\cal V}$ contains the bounded functions of $M^2$ and ${\cal W}$ is  generated by the operators $\zeta_0$ and 
\begin{equation}
K = i \frac{\partial}{\partial \zeta_0}.
\end{equation} 
Then ${\cal S}_0'$ is generated by the operators $\zeta_0$, $K$ and $M^2$, and it coincides with ${\cal U}^{(n)}$. The condition (\ref{Condition}) is satisfied and the Assumption \ref{Frame} is valid.

\end{document}